%% file: pairing2.tex
\documentclass[twocolumn,prb,aps,showpacs]{revtex4}
\usepackage{epsf,latexsym,graphicx}

\begin{document}

\input{def.tex}
\def \ba {\begin{eqnarray}}
\def \ea {\end{eqnarray}}
\def \vk {\mathbf{k}}
\def \vq {\mathbf{q}}
\def \Bi2212 {Bi$_2$Sr$_2$CaCu$_2$O$_{8+\delta}$}
\def \LSCO {La$_{2-x}$Sr$_x$CuO$_4$}
\def \YBCO {YBa$_2$Cu$_3$O$_{6.6}$}

\title{Comments on the $d$-wave pairing mechanism for cuprate high $T_c$ superconductors:
Higher is different? }

\author{Han-Yong Choi}
\affiliation{Department of Physics, Sung Kyun Kwan University, Suwon 440-746, Korea. \\
Asia Pacific Center for Theoretical Physics, Pohang 790-784,
Korea.} \email{hychoi@skku.ac.kr.}


\begin{abstract}

The question of pairing glue for the cuprate superconductors (SC)
is revisited and its determination through the angle resolved
photo-emission spectroscopy (ARPES) is discussed in detail. There
are two schools of thoughts about the pairing glue question: One
argues that superconductivity in the cuprates emerges out of
doping the spin singlet resonating valence bond (RVB) state.
Since singlet pairs are already formed in the RVB state there is
no need for additional boson glue to pair the electrons. The
other instead suggests that the $d$-wave pairs are mediated by
the collective bosons like the conventional low $T_c$ SC with the
alteration that the phonons are replaced by another kind of
bosons ranging from the antiferromagnetic (AF) to loop current
fluctuations. An approach to resolve this dispute is to determine
the frequency and momentum dependence of the diagonal and
off-diagonal self-energies directly from experiments like the
McMillan-Rowell procedure for the conventional SC. In that a
simple $d$-wave BCS theory describes superconducting properties of
the cuprates well, the Eliashberg analysis of well designed high
resolution experimental data will yield the crucial frequency and
momentum dependence of the self-energies. This line of approach
using ARPES are discussed in more detail in this review, and some
remaining problems are commented.

\end{abstract}

\maketitle


\section{Introduction}

After all the 26 years of unprecedentedly intense research into
the mysteries of the cuprate superconductors, we still have not
reached consensus on the mechanism of superconductivity.
Basically almost all conceivable tools have been applied to the
cuprates, which renders them arguably the best studied material
perhaps except for the semiconductors. Many anomalous properties
are abundant in the normal state above the superconducting
critical temperature, $T_c$, while the superconductivity itself
is well described by a generalization of the Bardeen, Cooper, and
Schrieffer (BCS) theory to the $d$-wave symmetry order parameter.
But, of course, that does not reveal the pairing mechanism. The
problem of understanding the cuprate superconductors (SC) is often
compared with that of conventional SC faced by BCS. However, these
problems are quite different in nature.

The main facts which the BCS theory tried to explain
were\cite{Bardeen57pr}: (1) a second-order phase transition at
$T_c$, (2) an electronic specific heat varying as $\exp(-T_0/T)$
near $T=0$ K and other evidence for an energy gap for individual
particle-like excitations, (3) the Meissner-Ochsenfeld effect
$(B=0)$, (4) effects associated with infinite conductivity
$(E=0)$, and (5) the dependence of $T_c$ on isotopic mass, $T_c
\sqrt{M}=\rm{const}$. All of these facts were explained by the
concept of phonon-mediated $s$-wave pairing of electrons of the
opposite spin and momentum and its generalization to many
electrons. For the cuprate superconductivity, the concept of
pairing is also valid,\cite{Gough87nature} and the orbital
symmetry is the $d$-wave one.\cite{Wollman93prl} What is not
established is what causes the pairing. Does higher $T_c$ imply a
different mechanism? For the conventional SC, all the normal (N)
state properties above $T_c$ are normal and describable in terms
of the Fermi liquid theory. All the facts BCS tried to explain as
summarized above were concerned with the transition between the N
and SC states.

On the other hand, for the cuprate superconductors, the
transition between N and SC states is describable in terms of the
$d$-wave BCS theory as well without revealing the pairing
mechanism, but, the non-superconducting states are anomalous.
Because the pairing is a Fermi surface instability in the normal
state, the pairing interaction should be manifest in other normal
state properties, that is, it should be able to account for the
normal state experimental observations unless nullified, for
example, by the symmetry. Therefore, the main requirements the
putative pairing interaction in the cuprates must explain are (1)
$d$-wave pairing of $T_c> 150$ K, (2) the anomalous normal state
near optimal doping, and (3) the pseudogap state of $T_c<T<T^*$,
where $T^*$ is the pseudogap temperature, and the transitions
between them. These must be understood in their totality; that is
the question.

Ideas proposed as a pairing mechanism may be classified into two
groups. One group of thoughts advocates that the
superconductivity in cuprates emerges upon doping out of the
resonating valence bond (RVB) singlet pairs formed by the
exchange coupling $J$.\cite{Anderson87science} Because the
singlet pairs are already formed in the RVB state, there is no
need for additional boson glue to pair the
electrons.\cite{Anderson07science} As charges are introduced
through doping, the singlet pairs become mobile and
superconductivity emerges naturally in the RVB scheme as an
optimal compromise of the competition between the kinetic energy
and singlet formation. On the contrary, the other school of
thoughts argues that the BCS pairing scheme with the mediating
boson works for the cuprates as well as the conventional SC. The
only difference is that the phonons in the low $T_c$ SC are
replaced by another kind of bosons. The suggested bosonic glue
ranges from the spin
fluctuations\cite{Monthoux07nature,Dahm09naturephys,Scalapino10physicac}
to the loop current
fluctuations.\cite{Varma06prb,Aji07prl,Aji10prb} Of course they
must satisfy the requirements above.

A way to resolve this dispute is to determine the frequency
dependence of the diagonal self-energy $\Sigma(\vk,\omega)$ and
off-diagonal self-energy $\phi(\vk,\omega)$ directly from
experiments like the McMillan-Rowell procedure for the
conventional SC.\cite{McMillan65prl} The cause of the frequency
dependence can provide an important clue about the pairing
mechanism.\cite{Maier08prl} The diagonal self-energy is also
called normal self-energy (``normal'' here means the
particle-hole channel and should not be confused with the
``normal'' as in the normal state meaning above $T_c$), and
off-diagonal self-energy is also called anomalous self-energy or
pairing self-energy. This approach should serve better than
numerical constructions and computations of the Hubbard-like
models.

Recall that a simple $d$-wave BCS model describes the SC
properties of the cuprates well. One can then expect that the
Eliashberg theory,\cite{Eliashberg60jetp} the extension of BCS
theory with dynamics built in, shall describe them better.
Therefore, the Eliashberg analysis of well designed high
resolution experimental data will yield the crucial frequency
dependence of the self-energies. This line of approach using the
angle resolved photo-emission spectroscopy (ARPES) are discussed
in more detail in this review. This is a generalization of the
McMillan-Rowell procedure to $d$-wave pairing
superconductors.\cite{McMillan65prl} For that, it is essential to
have the momentum resolved experimental inputs. They are provided
by the high resolution laser ARPES.

\section{pairing interaction}

There already exist many reviews on the pairing glue question. We
will only mention some more recent
reviews.\cite{Monthoux07nature,Abrahams11bcs,Lee08rpp,Scalapino10physicac,
Eschrig06aip,Choi11fop, Carbotte11rpp,Hackl10epjst,Zaanen10arXiv}
In a book commemorating the 50th anniversary of the BCS theory,
Abrahams wrote an excellent review on the evolution of the
theoretical ideas about the cuprate
superconductivity.\cite{Abrahams11bcs} He also gave a balanced
view on several ideas about the pairing mechanism including the
RVB, the antiferromagnetic (AF) spin fluctuations, and the loop
currents, among others. In Ref.\ \onlinecite{Norman11overview},
Norman gave a brief overview of the problem of high temperature
superonductivity in cuprates with an emphasis on theoretical
ideas. Lee focused on the development of the RVB theory and gave
some critical comments on other ideas in Ref.\
\onlinecite{Lee08rpp}. Scalapino made cases for the spin
fluctuations idea in Ref.\ \onlinecite{Scalapino10physicac}, and
Eschrig wrote a review on the experimental and theoretical works
on the interaction between single-particle excitations and
collective spin $S=1$ excitations in Ref.\
\onlinecite{Eschrig06aip}. Carbotte $et~al.$ mainly collected
experimental works in Ref.\ \onlinecite{Carbotte11rpp}. Hackl and
Hanke reported the current status of research and understanding
the high temperature superconductivity focusing on the cuprates
and pnictides in Ref.\ \onlinecite{Hackl10epjst}. Zaanen, in
Ref.\ \onlinecite{Zaanen10arXiv}, gave an overview of high
temperature superconductivity with a personal flavor.

In the spin fluctuation picture, the pairing is viewed as arising
from the exchange of particle-hole spin $S=1$ fluctuations whose
dynamics reflect the frequency spectrum seen in inelastic neutron
scattering (INS). In Ref.\ \onlinecite{Scalapino10physicac}, in
support of pairing via the spin fluctuations Scalapino notes the
commonalities among the heavy fermion, cuprate and Fe
superconductors, and argue that: (a) Their chemical and structural
makeup, their phase diagrams, and the observation of a neutron
scattering spin resonance in the superconducting phase support
the notion that they form a related class of superconducting
materials. (b) A number of their observed properties are described
by Hubbard-like models. (c) Numerical studies of the effective
pairing interaction in the Hubbard-like models find
unconventional pairing mediated by an $S = 1$ particle-hole
channel. He proposes that spin-fluctuation mediated pairing
provides the common thread which is responsible for
superconductivity in all of these materials. Although there is a
doubt whether the Hubbard type models contain high temperature
superconductivity,\cite{Aimi07jpsj} many groups subscribe to this
view.\cite{Monthoux07nature,Dahm09naturephys,Maier08prl,Kyung09prb,Eschrig06aip,Abanov03aip}

For instance, Dahm $et~al.$ analyzed the ARPES spectra of the
detwinned \YBCO\ single crystal and claimed that a self-consistent
description of the spectra can be obtained by modeling the
effective pairing interaction in terms of the spin susceptibility
$\chi(\vq,\w)$ determined by INS on the same single crystal. They
solved a simplified form of the $d$-wave Eliashberg equation by
neglecting the pairing dynamics (that is, the frequency
dependence of the pairing self-energy) and found that, in
addition to the successful description of the ARPES spectra, the
$T_c$ computed with this coupling constant exceeds 150 K,
demonstrating that spin fluctuations have sufficient strength to
mediate high-temperature superconductivity. See, however, the
discussion Sec.\ \ref{subsec:eliashberg-function} below for a
contrary view.

Kyung and his collaborators also reached a similar conclusion
about the nature of pair formation from the cellular dynamical
mean-field theory of the two-dimensional (2d) Hubbard
model.\cite{Kyung09prb} They found that the retardation effects
in the $d$-wave pairing of the 2d Hubbard model have the
corresponding energy scales with the short-range spin
fluctuations, and suggested that the low energy dynamics is
important for pairing.

Khatami $et~al$, on the other hand, investigated the spin
fluctuation approximation and reached a different
conclusion.\cite{Khatami09prb} They examined the validity of the
spin susceptibility glue approximation in a 2d Hubbard model for
cuprates using the dynamical cluster approximation with a quantum
Monte Carlo algorithm as a cluster solver. By comparing the
leading eigenvalues and corresponding eigenfunctions of the
dynamical cluster calculation and spin susceptibility
approximation pairing matrices, they found that the spin
susceptibility fails to capture the leading pairing symmetries
seen in the dynamical cluster approximation. Their results imply
that the low energy dynamics is less important than the high
energy contribution for pairing. This is in line with Anderson's
criticism of the glue idea.

Varma also belongs to the pairing glue camp. He and his
collaborators developed the idea that the pairing is mediated by
the local quantum critical fluctuations of the loop current
order.\cite{Varma06prb} As emphasized earlier in the Introduction,
the $d$-wave pairing, pseudogap, and anomalous normal state must
be understood as a whole. In this regard, it is important to note
that Varma and his collaborators have also demonstrated that the
loop current order is responsible for the pseudogap and its
fluctuations give rise to the marginal Fermi liquid
behavior\cite{Varma89prl} which underlies the anomalous normal
state above $T_c$. See Ref.\ \onlinecite{Varma06prb} for a
thorough review and Ref.\ \onlinecite{Choi11fop} for a recent
discussion of the idea in view of the ARPES analysis.

In dismissing the idea of the boson glue,\cite{Anderson07science}
Anderson argued that the singlet pairs are bound by the AF
superexchange coupling $J$ which is generated by virtual
excitations above the Mott gap set by the onsite repulsion $U$.
The pairing interaction is essentially instantaneous and there is
no low energy dynamics. That is, there is no pairing glue.

Of course, the ultimate question is: What does the experiment
tell us about the dynamics of the pairing interaction? Just as
the spatial structure of the pairing interaction can be
determined from the $\vk$-dependence of the superconducting gap,
the dynamics of the interaction is reflected in the frequency
dependence of the gap. This frequency structure of the gap is
reflected in a variety of experiments. For example, the analysis
of structure in ARPES\cite{Schachinger08prb} and infrared
conductivity\cite{Hwang07prb,Heumen09prb} have suggested that the
dynamics is determined by spin-fluctuations. As was observed
early on, the cuprate superconductivity is well described by the
simple $d$-wave BCS theory. Then the $d$-wave Eliashberg theory
may be utilized to extract the crucial dynamical information
about the cuprate superconducting properties. This path may be
pursued more systematically by tracking the angular dependence as
well as the frequency dependence. This is what we turn to now.

\section{dynamics extracted from ARPES}

This approach is being carried out especially thanks to the much
improved momentum and energy resolution of the laser ARPES. In
three recent papers,\cite{Bok10prb,Yun11prb,Zhang12prb} the
present author and his collaborators presented the results
extracted from the fits to the slightly underdoped Bi2212 with
$T_c=89$ K and the pseudogap temperature of $T^* \approx 160 $ K.
The analysis of other doping concentration crystals are also being
carried out.

\begin{figure}
\includegraphics[width=3.5in]{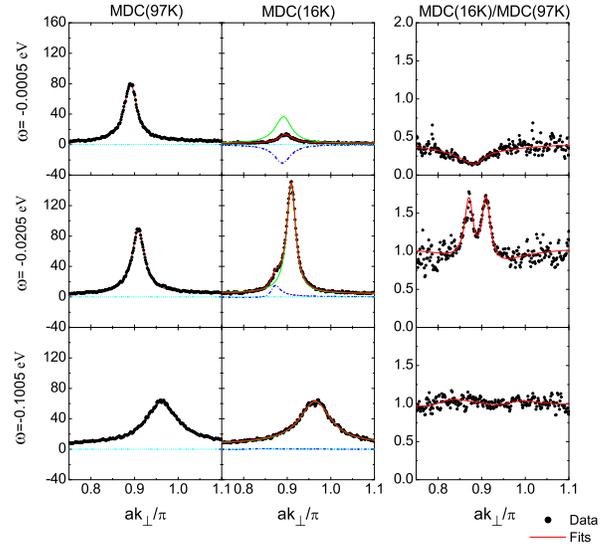}
\vspace{-0.2in}
 \caption{ The representative MDC as a function of
the momentum measured from the $(\pi,\pi)$ point along the tilt
angle $\th=20^{\circ}$. The dots are the experimental data and the
solid red lines are the fitting. The first and second columns
show the fitting in the normal state at $T=97$ and in the SC
state at $T=16$ K, respectively. The last column is the MDC ratios
of SC to normal states.}\label{MDCfitSC}
\end{figure}

The most crucial is the observation of the particle-hole mixing in
the momentum distribution curve (MDC) in SC state by the high
resolution laser ARPES which opened up a new window to probe the
fundamental physics of high temperature
SC.\cite{Zhang12prb,Yun11prb} See the middle row of Fig.\ 1 which
shows the MDC along the cut $\th=20^{\circ}$. We use the angle
$\th$ from the diagonal direction to label each cut in the
Brillouin zone as shown in Fig.\ 2. The actual momentum paths are
shown by the slightly curved thick bars along each cut. The
$k_\perp$ is the distance from the $(\pi,\pi)$ point. The first
and second columns of the Fig.\ 1 show the normal state at $T=97$
and SC state at $T=16$ K, respectively. For the second column,
the red, green, and blue represent the total, particle, and hole
contributions. Notice that in addition to the main peak near
$k_\perp a/\pi\approx 0.91 $ from the original quasi-particle
branch, there exists the secondary peak at $k_\perp a/\pi\approx
0.87 $ from the hole branch. This is a direct observation of the
particle-hole mixing deep in the SC state and can be utilized to
obtain the crucial frequency dependence of the self-energy of the
cuprates.

\begin{figure}
\includegraphics[scale=0.25]{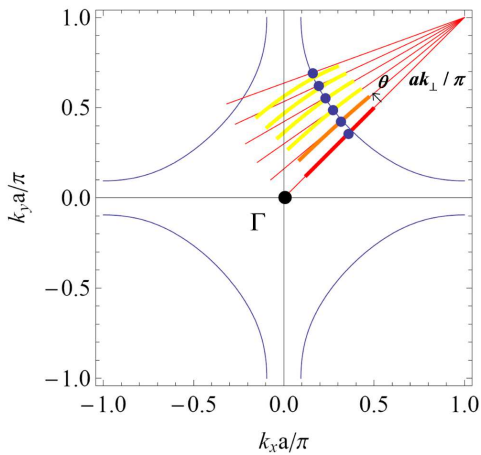}
\caption{The Fermi surface of Bi2212 in the first Brillouin zone.
The blue solid curve is a calculated Fermi-surface and the solid
dots are experimentally determined FS at $\theta=0$, 5, 10, 15,
20, and 25 degrees. $k_\perp$ is the distance from the
$(\pi,\pi)$ point. The thick bars along each cut indicate the
ranges of experimentally measured ARPES MDC data.} \label{fig:FS}
\end{figure}

Another crucial point is that the Eliashberg analysis of the
ARPES data can distinguish the Eliashberg functions in the
diagonal and off-diagonal (pairing) channels, $\a^2
F^{(+)}(\vk,\vk',\w)$ and $\a^2 F^{(-)}(\vk,\vk',\w)$,
respectively. See Eq.\ (\ref{a2f}) below. To our knowledge, this
separation can only be accomplished from analysis of ARPES data.
This is particularly interesting because it offers a way to
disentangle the boson spectrum. As Anderson
commented,\cite{Anderson07science} the strong onsite repulsion
$U$ causes the broad structure in the electrons' energy
distribution functions. This may naively be described by coupling
to a broad boson spectrum which, however, doesn't help with pair
binding. In the Eliashberg framework, $\a^2 F^{(+)}(\vk,\vk',\w)$
and $\a^2 F^{(-)}(\vk,\vk',\w)$ represent, respectively, the boson
spectrum which electrons are coupled to and that which helps with
pair binding.

This is an extension of the tunneling experiments and analysis
with which it was definitely established that the pairing in
metals like Pb is through exchange of
phonons.\cite{McMillan65prl} It should be remembered that to get
reliable information, it was necessary to have measurements of
conductance at different temperatures and range of voltages of
the order of the cut-off energy in the phonon spectrum to an
accuracy of 0.2 \%. Since the cut-off is an order of magnitude
higher and the angle-dependence of the spectra is crucial for the
cuprates, the demands on the quality of the data are only being
recently met through ultra-high resolution and stability of laser
based ARPES.

\subsection{Deduced diagonal self-energy}
\label{subsec:self-energy}

\begin{figure}
\includegraphics[scale=0.35]{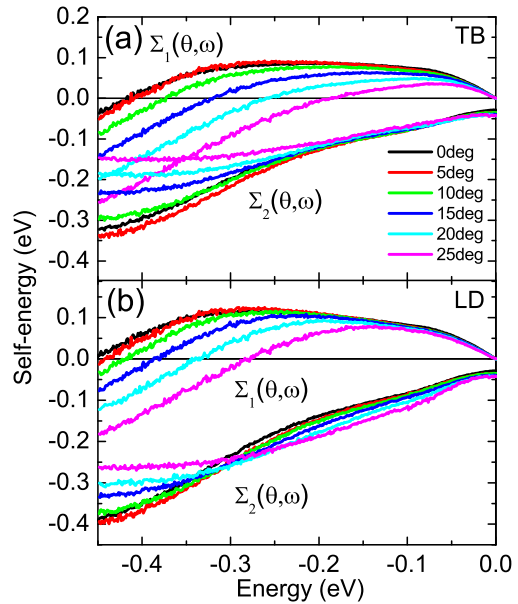}
\vspace{-0.4in}
 \caption{The real $\Sigma_1$ and  imaginary
$\Sigma_2$ parts of self-energy at $T = 107 K$  as a function of
energy $\omega$ for the tilt angles $\theta$ = 0, 5, 10, 15, 20,
and 25 degrees with respect to the diagonal cut in the Brillouin
zone. Plot (a) is the results obtained using a tight-binding
dispersion $\xi(\vk)$ while (b) is given to compare the results
if a linear extrapolation of the band-structure from that near
the Fermi energy is adopted.} \label{self-energy}
 \end{figure}

The quantity that we will focus is the angle and frequency
dependence of the diagonal self-energy $\Sigma(\theta,\omega)$.
We refer to the papers Ref.\
\onlinecite{Bok10prb,Yun11prb,Zhang12prb} for the detailed
procedure for inverting ARPES. A very similar approach using
ARPES only along the nodal direction was reported in Ref.\
\onlinecite{Schachinger08prb}. Here we only provide a summary of
the results. We were able to deduce the angle and frequency
dependence of the normal state self-energy
$\Sigma(\theta,\omega)$ at $T= 107$ K. There is no perceptible
dependence on $|{\vk}-{\vk}_F|$ as can be seen from the first
column of Fig.\ \ref{MDCfitSC} in that the normal state MDC are
almost perfect Lorentzian. The obtained self-energies are shown
in Fig.\ \ref{self-energy}. The subscripts 1 and 2 stand for the
real and imaginary parts, respectively. The plot
\ref{self-energy}(a) is the results obtained using a tight-binding
dispersion $\xi(\vk)$, while \ref{self-energy}(b) is obtained
using a linear dispersion. It is given for comparison to check the
effects of different bare dispersions. Here the bare dispersion
$\xi(\vk)$ actually represents the renormalized dispersion but
without including the effects of the putative interaction
$\alpha^2 F$.

Two points are to be noticed about the self-energy at $T=107 $ K
in the normal state: (1) In the low energy regime $(-0.05 \lesssim
\omega <0)$, the $\Sigma_1(\theta,\omega)$ is almost angle
independent, and (2) the zero crossing energy $\omega_0=|\omega|$
of $ \Sigma_1(\theta,\omega)$, that is,
$\Sigma_1(\theta,-\omega_0)=0$, decreases monotonically as the
angle increases. From $\omega_0\approx 0.4$ at $\theta=0$ to
$\omega_0\approx 0.2$ eV at $\theta=25^\circ$. In order to
understand these results in terms of the effective interaction
$\alpha^2 F$, recall that the $\Sigma_1$ is the shift of the
renormalized dispersion from the bare one and is proportional to
the $\alpha^2 F$ within the Eliashberg framework. Also recall
that the zero of the real part of a causal function corresponds
to a peak (or, saturation) of the imaginary part which in turn
corresponds to the cutoff of $\alpha^2 F$. Then the point (1)
implies that the $\alpha^2 F$ is almost angle independent below
$~0.05$ eV, and (2) implies that the cutoff energy of the
Eliashberg function decreases as the angle increases. These are
indeed what we found by the maximum entropy method to invert the
$d$-wave Eliashberg equation as given in Eq.\ (\ref{Eliad}). See
the Fig.\ \ref{normal F} below. This will be discussed in the next
section.

\begin{figure}
\includegraphics[scale=0.35]{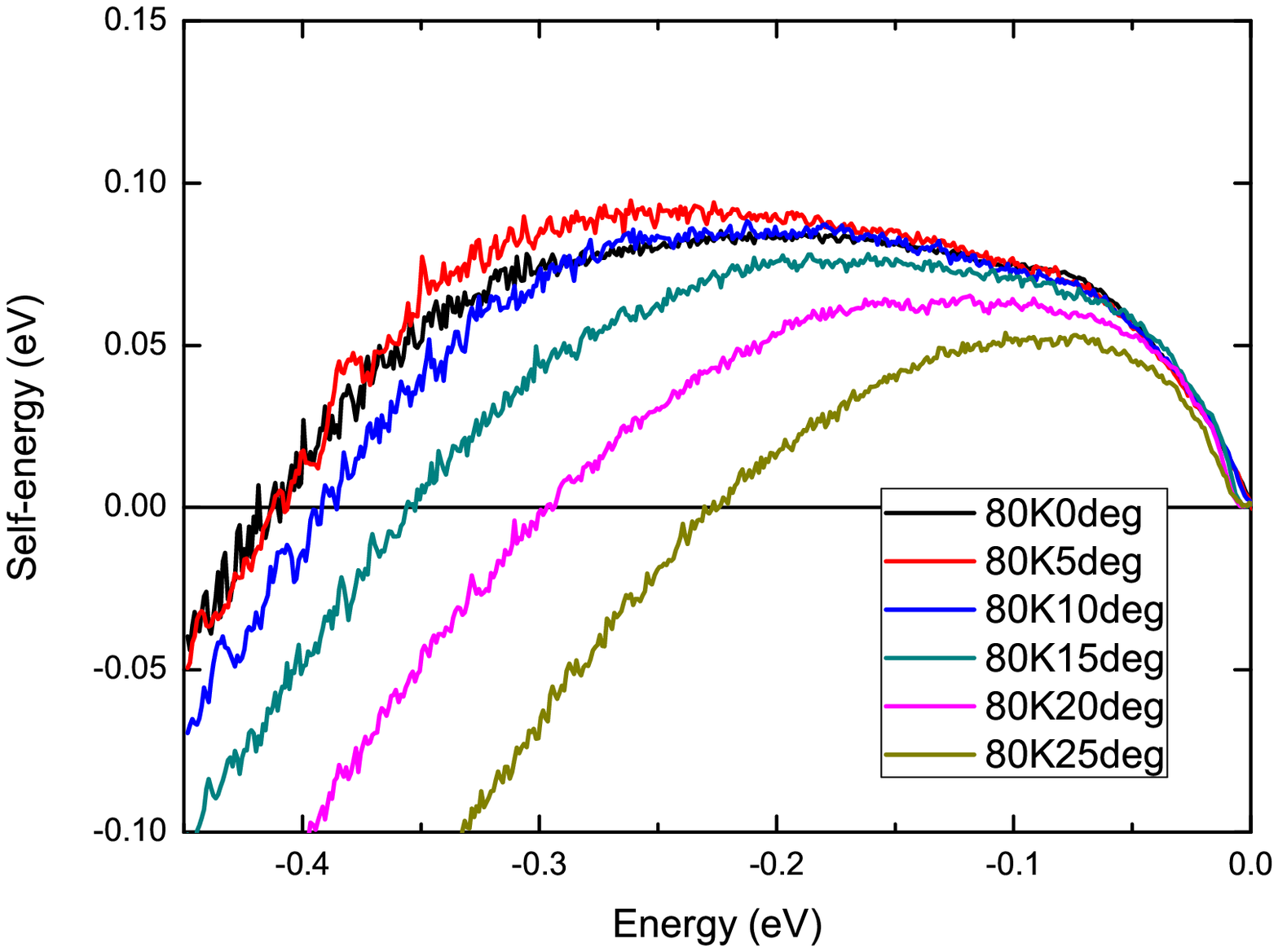}
\includegraphics[scale=0.35]{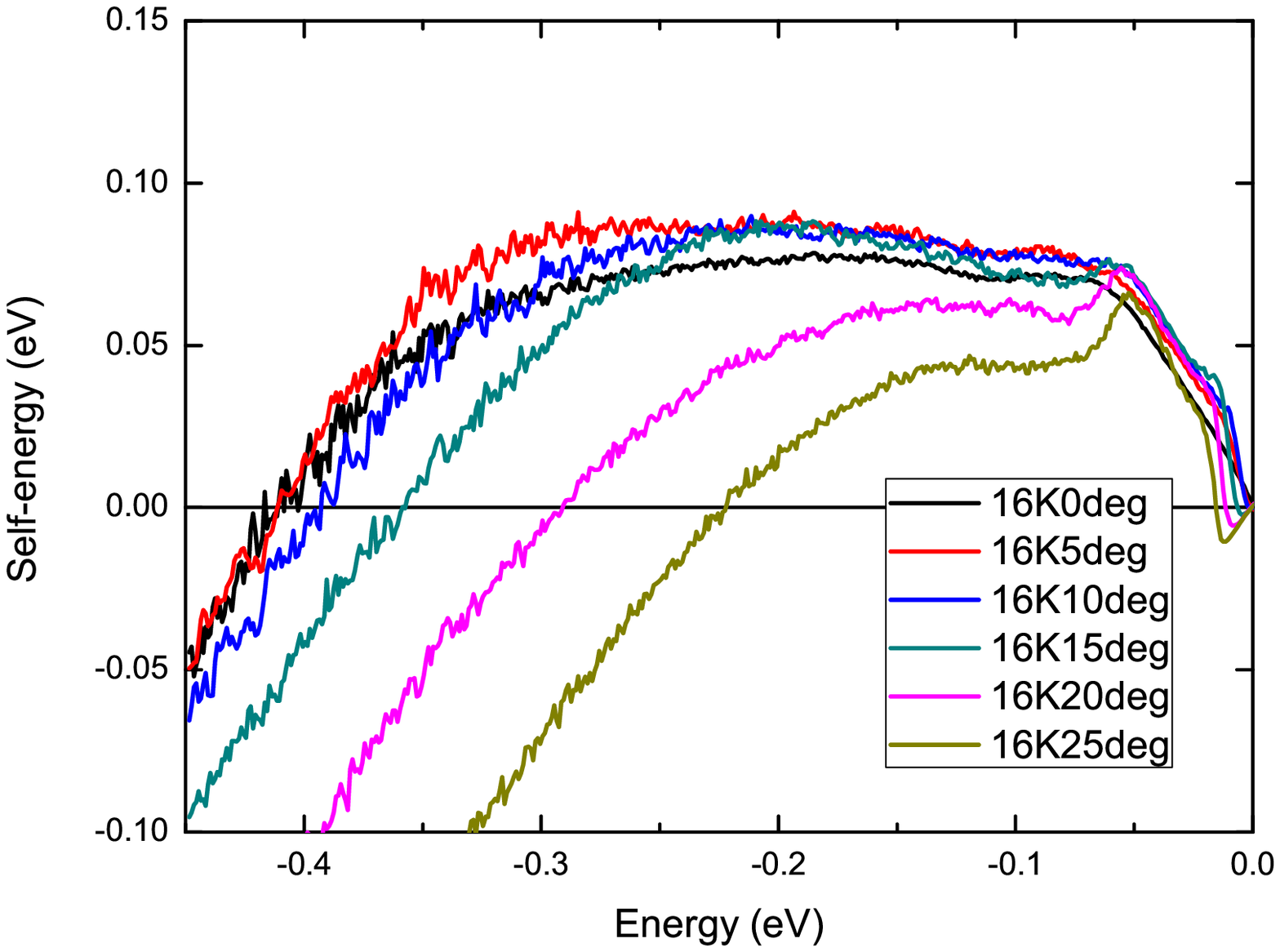}
\caption{The real part of the self-energy. (a) is at $T=80$ K
slightly below $T_c$, and (b) is at $T=16$ K deep in the SC
state. Notice that the two observations made at $T=107$ K are
also valid at $T=80$ K; the overlap of the curves for small $\w
\gtrsim -0.05$ eV and the decreasing cutoff energy as a function
of the angle. Only in the low temperature the SC induced features
show up near $-0.02$ and $-0.05$ eV as can be seen from the plot
(b). } \label{realsigma}
\end{figure}

We now consider how the angle and frequency dependence of the
self-energy vary in the SC state. The real part of the
self-energy is shown in Fig.\ \ref{realsigma}. Plot
\ref{realsigma}(a) is at $T=80$ K just below $T_c=89$ K, and (b)
is deep in the SC state at $T=16$ K. At $T=80$ K, above two
observations at $T=107$ K remain valid, that is, the angle
independence of $\Sigma_1$ in the low energy and decreasing zero
crossing energy as the angle increases from the nodal cut. Only
deep in the SC state at $T=16$ K as shown in Fig.\
\ref{realsigma}(b) the SC induced changes in the $\Sigma_1$ show
up. Two structures emerge near $\omega\approx -0.02$ and $-0.05$
eV. Both are consistent with the $d$-wave pairing gap. The two
structures imply that the $\a^2F$ will exhibit two additional
features to that in the normal state. To that we turn now.

\subsection{Eliashberg function}
\label{subsec:eliashberg-function}

While the self-energy extraction does not need an underlying
theory except for the observation that cuprate superconductivity
follows the $d$-wave BCS theory, the description of the
self-energy in terms of boson spectrum requires that the
Eliashberg-type theory is valid for the cuprates. Although it is
not yet settled if the Eliashberg formalism is valid for the
cuprates (see below in Sec.\ V), we will use it to discuss the
boson spectrum. Please see Ref.\ \onlinecite{Yun11prb} for the
results at various temperatures and angles.

The Eliashberg equation is given by
 \ba
\widetilde{\Sigma}(\mathbf{k},\omega)=\int^{\infty}_{-\infty}
d\epsilon \int^{\infty}_{-\infty} d\epsilon'
\frac{f(\epsilon)+n(-\epsilon')}{\epsilon+\epsilon'-\omega-i\delta} \nonumber\\
 \times \sum_{\mathbf{k'}}A_{S}(\mathbf{k'},\epsilon)\alpha^{2}F^{(+)}(\mathbf{k},\mathbf{k'},\epsilon'),
 \nonumber\\
 X(\mathbf{k},\omega)=\int^{\infty}_{-\infty}d\epsilon \int^{\infty}_{-\infty}d\epsilon'
\frac{f(\epsilon)+n(-\epsilon')}{\epsilon+\epsilon'-\omega-i\delta}
\nonumber\\
 \times \sum_{\mathbf{k'}} A_{X}(\mathbf{k'},\epsilon)
\alpha^{2}F^{(+)}(\mathbf{k},\mathbf{k'},\epsilon'),
 \nonumber \\
 \phi(\mathbf{k},\omega)=\int^{\infty}_{-\infty}d\epsilon \int^{\infty}_{-\infty}d\epsilon'
\frac{f(\epsilon)+n(-\epsilon')}{\epsilon+\epsilon'-\omega-i\delta}
\nonumber\\
 \times \sum_{\mathbf{k'} }A_{\phi}(\mathbf{k'},\epsilon)
\alpha^{2}F^{(-)}(\mathbf{k},\mathbf{k'},\epsilon'), \label{Eliad}
 \ea
where $f$ and $n$ are the Fermi and Bose distribution functions,
respectively.
 \ba
\Sigma(\vk,\w)=\widetilde{\Sigma}(\vk,\w) +X(\vk,\w)
 \ea
is the diagonal self-energy and $\phi(\vk,\w)$ is the off-diagonal
self-energy. The spectral functions are given by
 \ba
A_S (\vk,\epsilon)&=& -\frac{1}{\pi} Im
\frac{W}{W^2-Y^2-\phi^2}=A_S(\vk,-\epsilon),
 \nonumber \\
A_X (\vk,\epsilon)&=& -\frac{1}{\pi} Im \frac{Y}{W^2-Y^2-\phi^2}=
 -A_X(\vk,-\epsilon),
 \nonumber \\
A_\phi (\vk,\epsilon)&=& -\frac{1}{\pi} Im
\frac{\phi}{W^2-Y^2-\phi^2}=
 -A_\phi(\vk,-\epsilon),
 \nonumber \\
A (\vk,\epsilon)&=& A_S (\vk,\epsilon)+A_X (\vk,\epsilon).
 \ea
We use
 \ba
W(\vk,\w)= \w -\widetilde{\Sigma}(\vk,\w), \nonumber \\
Y(\vk,\w)= \xi(\vk) +X(\vk,\w).
 \ea
The diagonal and off-diagonal Eliashberg functions are given by
 \ba
 \alpha^{2}F^{(+)}(\mathbf{k},\mathbf{k'},\epsilon') =
 \alpha_{ch}^{2}F_{ch}(\mathbf{k},\mathbf{k'},\epsilon')
 +\alpha_{sp}^{2}F_{sp}(\mathbf{k},\mathbf{k'},\epsilon'),
\nonumber \\
 \alpha^{2}F^{(-)}(\mathbf{k},\mathbf{k'},\epsilon') =
\alpha_{ch}^{2}F_{ch}(\mathbf{k},\mathbf{k'},\epsilon')
-\alpha_{sp}^{2}F_{sp}(\mathbf{k},\mathbf{k'},\epsilon'),
 \nonumber \\
\alpha^2 F^{(\pm)} (\th,\e')\equiv \left\langle
\f{\a^{2}(\th,\th')}{v_{F}(\th')}F^{(\pm)}(\th,\th',\e')
\right\rangle_{\th'}, \label{a2f}
 \ea
where $v_{F}(\th')$ is the angle dependent Fermi velocity and the
bracket implies the angular average over $\th'$. The subscripts
$ch$ and $sp$ represent, respectively, the time-reversal symmetry
conserving and breaking interactions. For example, to the latter
(former) belong the spin (charge) and current interactions. For
more technical details, please refer to the references Ref.\
\onlinecite{Yun11prb,Hong12unp}.

\begin{figure}
\includegraphics[scale=0.9]{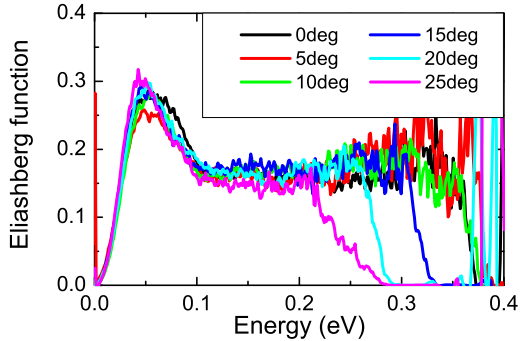}
\caption{The diagonal Eliashberg function $\alpha^2
F^{(+)}(\theta, \omega)$ at $T=107$ K extracted from the
self-energy shown in Fig.\ \ref{self-energy}. The curves overlap
below $0.02$ eV and the cutoff energy decreases as the angle
increases from the nodal cut.} \label{normal F}
\end{figure}

Using the deduced self-energy, we have inverted the $d$-wave
Eliashberg equation for the normal state self-energy to deduce the
normal Eliashberg function $\alpha^2 F^{(+)} (\theta, \epsilon)$.
The results are shown in Fig.\ \ref{normal F}. The deduced
$\alpha^2 F^{(+)} (\theta, \epsilon)$ behaves as expected before
in Section \ref{subsec:self-energy}. It is independent of angle
to an accuracy of about 10\% below an energy of about 0.2 eV.
Above this energy there is an angle dependent cutoff $\omega_c
(\theta)$. $\w_c$ decreases as the angle increases from $\sim 0.4$
eV at $\th=0$ to $\sim 0.2$ eV at $\th=25$ degrees. That is, the
only angle dependence of the Eliashberg function in the normal
state is the cutoff $\w_c(\th)$.

Now, we consider the diagonal Eliashberg function in the SC
state.\cite{Yun11prb} The principal conclusions are that along
the nodal cut ($\theta = 0$) the fluctuations below $T_c$, within
the uncertainty of determination, are almost unchanged from the
fluctuations above $T_c$. For larger energies above about 0.1 eV,
they are similarly unchanged from the fluctuations above $T_c$.
On the other hand, there is a growth of the peak around 50 meV
for the lower temperatures and larger angles, and the emergence
of a new peak at about $10-15$ meV. These features are probably
related to the loss of dissipation due to the opening of the
superconducting gap and would be interesting to study
theoretically in greater detail.

The deduced $\a^2 F^{(+)}(\th,\w)$ discussed above are consistent
with an earlier deduction \cite{Schachinger08prb} from ARPES
spectrum in the same compound, and also qualitatively consistent
with the deduction from optical conductivity
spectrum\cite{Hwang07prb,Heumen09prb}, which preferentially
weights the nodal quasi-particles because of their larger Fermi
velocity.

Now, let us consider the Eliashberg function along the pairing
channel, $\a^2 F^{(-)}(\th,\w)$. The information on the pairing
self-energy is contained only in the difference in the ARPES
spectra in the superconducting state and the normal state. This
difference is expected\cite{Yun11prb} to be less than 1\% above
an energy of a few times the superconducting gap. Above such
energy, the noise in the data is at present significantly larger
than 1\%. Therefore, we have not been able to extract the pairing
self-energy and to directly deduce $\alpha^2 F^{(-)}(\omega)$
from the data over the full energy range of 0.4 eV. But it would
be ideal to have data which is about 1/2 an order of magnitude
better to completely settle the shape of $\alpha^2
F^{(-)}(\theta, \epsilon)$.

Determination of the off-diagonal Eliashberg function $\alpha^2
F^{(-)}(\theta, \epsilon)$ offers perhaps the best way to decide
among the proposed ideas. In the loop current idea the $\alpha^2
F^{(+)}(\theta, \epsilon)$ and $\alpha^2 F^{(-)}(\theta,
\epsilon)$ have the same frequency dependence, while they are
different in the spin fluctuation scenario.

\section{Discussion of Results}

In this section, we discuss the implications of the finding that
the Eliashberg function $\alpha^2F^{(+)}(\theta, \omega)$ is
nearly independent of $\theta$ in the normal state and just below
$T_c$.\cite{Choi11fop} It should be remembered again that the
superconducting $T_c$ is the property of the normal states.
Indeed the normal state self-energy just above $T_c$ includes all
forms of scattering, both spin-dependent as well as
spin-independent scatterings from all initial ${\mathbf k}$ to
all final momenta ${\mathbf k'}$. Given this, the crucial issue
to address is how fluctuations which lead to a nearly
$\theta$-independent $\alpha^2F(\theta,\omega)$ are reconcilable
with the same fluctuations promoting the $d$-wave pairing.

The inadequacy of the AF fluctuations in this regard was given
previously.\cite{Choi11fop} Although the pairing Eliashberg
function has not been determined separately, the deduced diagonal
Eliashberg function alone is enough to reach this conclusion. The
idea is that in the AF spin fluctuation scenario the spin
susceptibility $\chi(\vq,\w)$ provides both $\a^2 F^{(\pm)}$: the
$s$- and $d$-wave projection of the spin susceptibility yield
$\a^2 F^{(+)}$ and $\a^2 F^{(-)}$, respectively. The nearly $\th$
independent $\a^2 F^{(+)}$ implies a small correlation length
$\xi/a \lesssim 1$, where $a$ is the lattice constant. The INS
results on YBaCuO indeed found a small AF correlation length
$\xi/a \lesssim 1$ near optimal doping
concentration.\cite{Balatsky99prl} The small correlation length
means a small $d$-wave projection in the AF spin fluctuation
scheme. Then, if one interprets the deduced $\a^2 F^{(+)}(\th,\w)$
in terms of the AF fluctuations, s/he can not escape from the
conclusion of a small $d$-wave projection component and the low
$T_c$ smaller than 10 K.

The prospect of the loop current fluctuations was also discussed
in Ref.\ \onlinecite{Choi11fop}. The idea is that although the
fluctuation spectrum is nearly momentum independent, the coupling
vertex has a sufficiently large $d$-wave component in the loop
current fluctuations scenario.\cite{Aji10prb} Here we will add
arguments against the AF fluctuation scenario deep in the SC
state.

\begin{figure}
\includegraphics[scale=0.4]{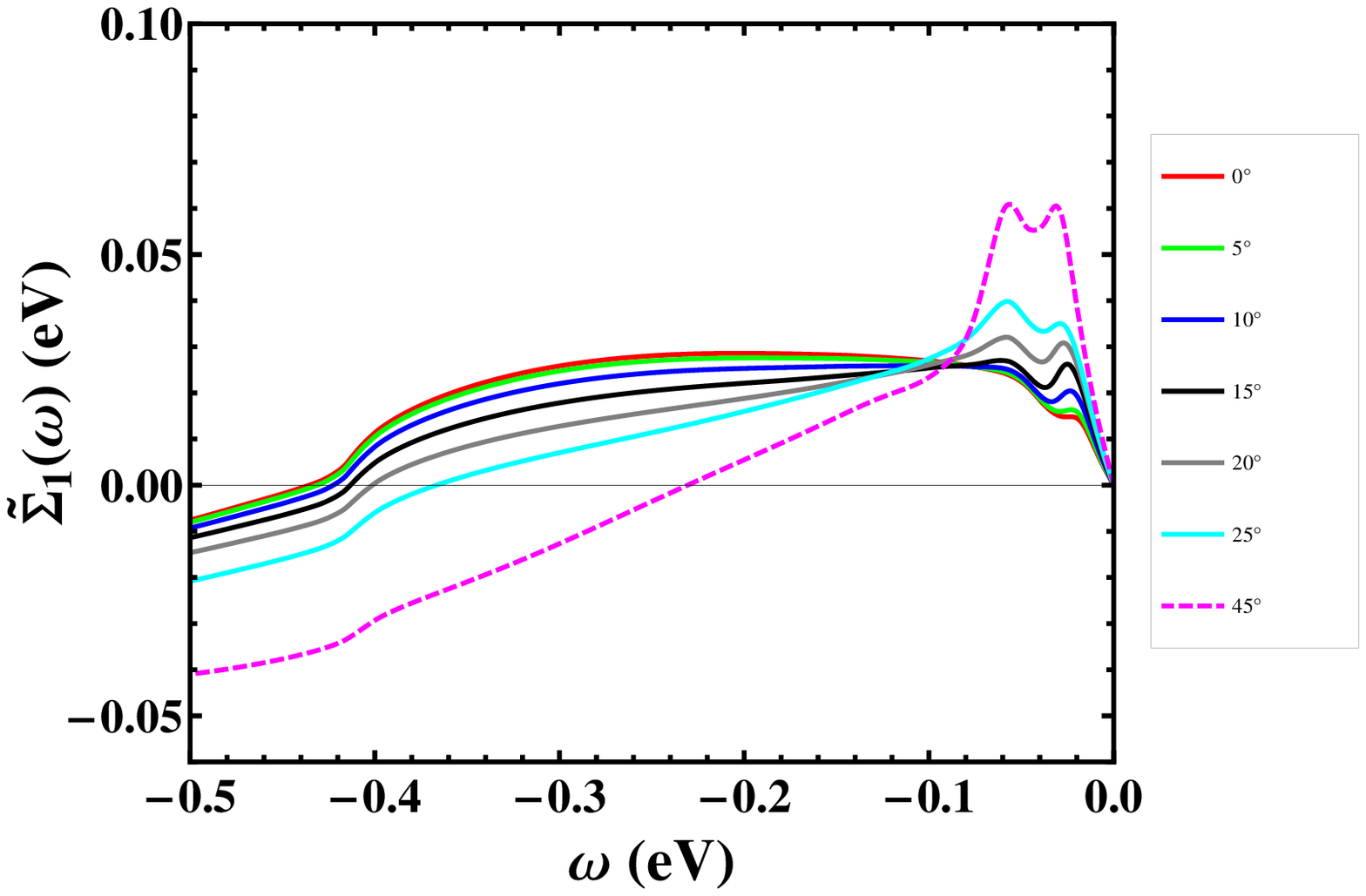}
\includegraphics[scale=0.4]{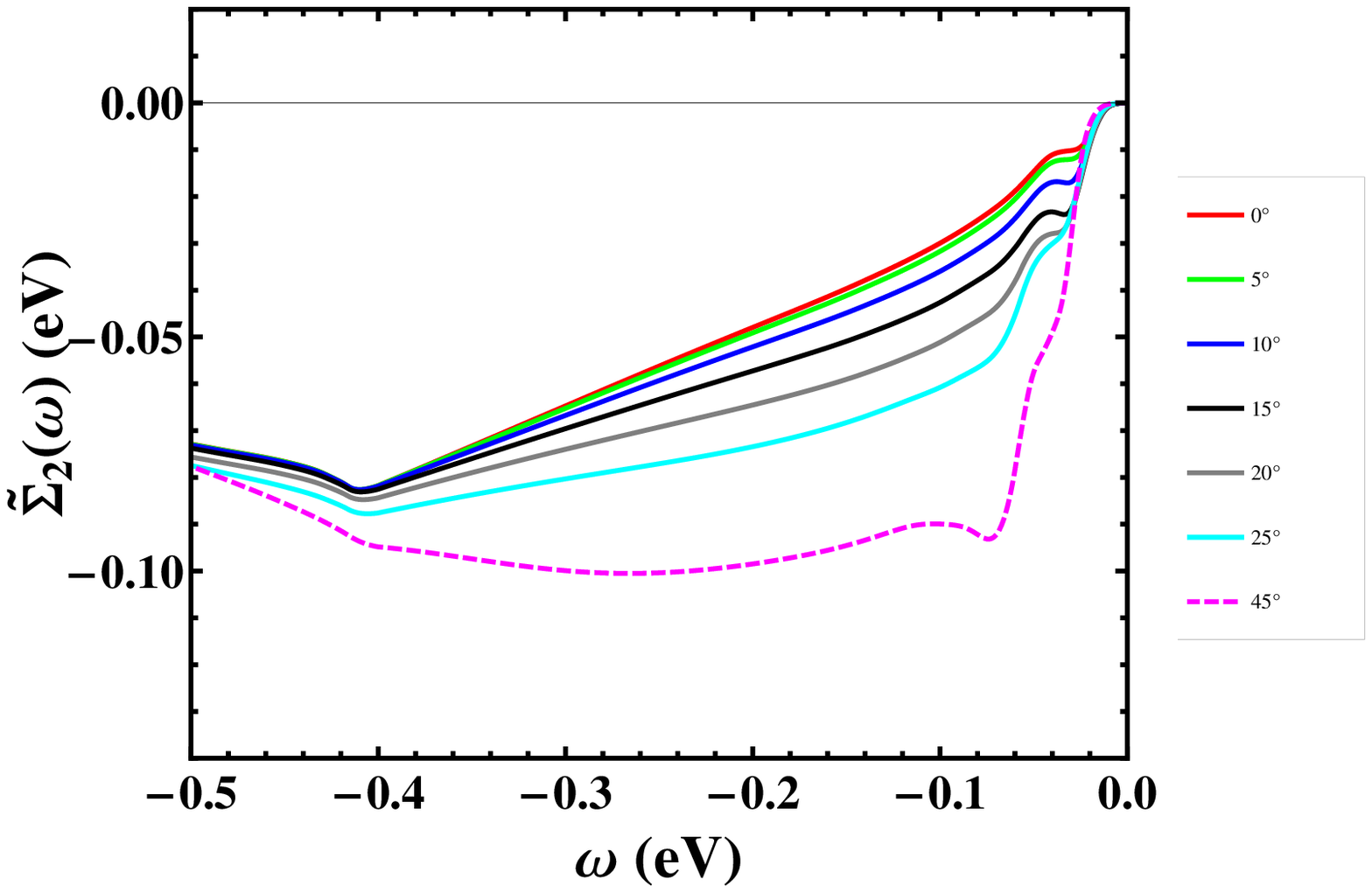}
\caption{The symmetric part of the self-energy computed from the
Vignolle spectrum of \LSCO. (a) is the real part and (b) is the
imaginary part. One should compare the plot (a) with the Fig.\
\ref{realsigma}(b). See the text for more details. }
\label{vignolle}
\end{figure}

The extracted real part of the diagonal self-energy at $T=16$ K
is shown in Fig.\ \ref{realsigma}(b). Two observations were made:
(1) almost angle independent $\Sigma_1(\th,\w)$ in the low energy
except for the SC induced features near $\omega\approx -0.02$ and
$-0.05$ eV, (2) decreasing zero crossing energy $\omega_0$ as a
function of the angle. To check if the two observations can be
explained by the AF spin fluctuation idea, we chose the simplest
system where the magnetic spectrum is well known. We took the
Vignolle spectrum of single layer \LSCO \ with $x=0.16$ from
INS\cite{Vignolle07naturephys} for the $\alpha^2 F_{sp}$ and
$\alpha^2 F_{ch}=0$ for Eq.\ (\ref{a2f}). Although the ARPES
results were extracted from the Bi2212, we believe in the
commonality of the cuprates phenomenology and trust that the
deduced ARPES can certainly be compared with the INS results.

These were solved self-consistently without assuming a separable
$\vk,\w$ form or particular momentum dependence of the pairing
self-energy $\phi(\vk,\w)$. Details are reported in Ref.\
\onlinecite{Hong12unp}. The symmetric part of the diagonal
self-energies are shown in Fig.\ \ref{vignolle}. Plots
\ref{vignolle}(a) and (b) are the real and imaginary parts,
respectively. Clearly the peak features of $\widetilde{\Sigma}_1$
around $30-70$ meV go together with the decreasing zero crossing
energy $\w_0$ as a function of the angle. It is because the sharp
$\a^2 F$ peak at $(\pi,\pi)$ is most effective in scattering the
particles near the antinodal region ($\th\approx 45^\circ$). The
effects of the scatterings off the $(\pi,\pi)$ peak show up as a
fast increase of $-\Sigma_2(\th\approx45,\w)$ as $-\w$ increases.
$-\Sigma_2(\w)$ then makes a shallow peak and begins to decrease.
Recall that to a peak in the imaginary part of a causal function
corresponds a zero crossing of the real part of that function.
This is the reason why the peak features around $30-70$ meV go
together with the decreasing zero crossing energy in the AF spin
fluctuations scenario. As shown in Fig.\ \ref{vignolle}(a), the
peak feature around $30-70$ meV is too strong and the decrease of
the zero crossing energy is too slow compared with the ARPES
results shown in Fig.\ \ref{realsigma}(b).

The previous conclusion from the Ref.\ \onlinecite{Choi11fop} was
that if the fluctuations coupling to fermions revealed in ARPES
experiments were AF fluctuations, $T_c$ would have been less than
10 K. The reason is that if the correlation length of
fluctuations is not much larger than the lattice constant, the
projection to $d$-wave scattering is small compared to the
average or $s$-wave scattering. Hence, follows a very small
$T_c$. The SC state analysis briefed here reinforces the previous
conclusion from the normal state analysis that it is difficult to
understand the results of the ARPES analysis within the AF spin
fluctuations idea.

This view is in stark contrast with the result reported by Dahm
$et~al.$\cite{Dahm09naturephys} As discussed before in Sec.\ II,
they analyzed the charge- and spin-excitation spectra determined
by the ARPES and inelastic neutron scatterings on the same
crystal of \YBCO\ and claimed that a self-consistent description
of both spectra can be obtained by adjusting a single parameter,
the spin-fermion coupling constant. They suggested that the spin
fluctuations have a sufficient strength to mediate
high-temperature superconductivity.

We suggest that there is a more stringent test on the spin
fluctuation scenario. Because the commensurate and incommensurate
spin fluctuations of the Vignolle spectrum have sharp peaks, of
about $\xi/a\approx 2-3$ and $6-7$ in the low temperature limit,
their scatterings are selective in the momentum space. That is
why the peak of the real part self-energy shown in Fig.\
\ref{vignolle}(a) exhibits the strong angle dependence. This,
however, does not agree with what was deduced from the ARPES
analysis. The detailed frequency and angle dependence of the
self-energy need to be performed for the AF spin fluctuations
scenario before claiming it as the pairing interaction.

\section{outlooks}

In hindsight, the remarkable success of the BCS theory owes much
to the smallness of the ratio of the phonon energy to the Fermi
energy, $\hbar\w_D/\epsilon_F \ll 1$. Assured by the Migdal
theorem, it enables one to do the controlled perturbation
expansion. Moreover, this small ratio also tames the Coulomb
repulsion into the benign Coulomb pseudo-potential, $\mu^*$,
while at the same time this retardation effect limits the $T_c$.

For pairing due to the electron-electron interaction, one can
certainly try to invoke the Migdal theorem thanks to the small
ratio of the collective energy like the spin fluctuation energy
$\w_{sf}$ to the Fermi energy $\epsilon_F$. But there is no
rigorous grounds for it.\cite{Abanov03aip,Hertz76ssc} The
difficulty comes from the fact that in the electron system the
collective degrees of freedom like the spin fluctuations are
composed of the very same electrons that are being paired. One
should come up with a scheme to deal with this situation more
systematically.

The determination of the frequency dependence of pairing was
motivated by the idea that it could differentiate among the
proposed theories. The proponents for the RVB theory argue that
the pairing in the cuprates does not have the low energy dynamics.
On the other hand, the other group for the pairing glue like the
spin fluctuations or loop current fluctuations argues that the
pairing interaction has the low energy dynamics set by the
relevant fluctuations. However, even in the RVB theory the claim
of no low energy dynamics is not accepted unanimously. Attempts
to go beyond the mean-field RVB approach typically invoke the
gauge fluctuations as a way to enforce the no double occupancy
which generates significant low energy dynamics.\cite{Lee06rmp}
But, this dynamics of gauge fluctuations is difficult to compute
to compare with experiments quantitatively.

In this regard, two more features of the ARPES analysis will be
valuable: The angle dependence of the self-energy and the separate
determination of the two Eliashberg functions, the diagonal and
off-diagonal $\a^2 F^{(\pm)}$. The angle dependence of the
self-energy can be a stringent test as we argued in the previous
section. Also the possibility of independent extraction of $\a^2
F^{(+)}$ and $\a^2F^{(-)}$ from the diagonal and off-diagonal
self-energies, $\Sigma$ and $\phi$, respectively, is a unique
advantage of the ARPES analysis. As we stand now, determination
of $\phi(\vk,\w)$ suffers from loss of accuracy. As we commented
before, the information on the pairing self-energy is contained
in the difference of the ARPES intensities between the normal and
SC states. It would be ideal to have ARPES data of about half an
order of magnitude better to fix the pairing Eliashberg function
$\a^2 F^{(-)}$. This perhaps gives the best way to decide among
the proposed ideas.

Differences among the ideas should also show up in the excitations
of the proposed states. The elementary excitations in the RVB
state have reversed charge-statistics relations: They are neutral
spin-1/2 fermions and charge $\pm e$ spinless bosons, analogous
to the solitons in polyacetylene.\cite{Kivelson87prb} In the spin
fluctuations idea they are the usual fermions and $S_z=\pm 1$
spins. The excitations in the loop current state are also being
explored.\cite{Li12naturephys}

The phase diagram also offers a way to differentiate among
proposals. The RVB theory predicted a phase diagram where the
pseudogap temperature $T^*$ and SC critical temperature $T_c$
lines merge together as the doping concentration increases. This
is in contrast to the phase diagram of the pairing glue camp where
the $T^*$ and $T_c$ lines cross each other and $T^*$ line
continues inside the $T_c$ dome. In both scenarios the nature of
the pseudogap phase below $T^*$ determines the origin of the
pairing. In this regard, the nature of the pseudogap is the key.
And again, pseudogap, anomalous normal state, and
superconductivity should be understood in their totality. That is
the question.

\begin{acknowledgments}

The author would like to thank Jae Hyun Yun, Jin Mo Bok, Seung
Hwan Hong, Wentao Zhang, Prof.\ Xingjiang Zhou, and Prof.\ Chandra
Varma for the collaborations and useful comments. He also wishes
to express thanks to Prof.\ Dong Ho Kim for his invitation of this
article to the Journal of Korean Physical Society. This work was
supported by National Research Foundation (NRF) of Korea through
Grant No.\ NRF 2011-0005035.

\end{acknowledgments}

\bibliographystyle{naturemag}

\bibliography{review}

\end{document}

%% file: def.tex
\newcommand{\av}[1]     {\langle #1 \rangle}
\newcommand{\Av}[1]     {\left\langle #1 \right\rangle}
\newcommand{\comm}[1]   {\left[ #1 \right]}
\newcommand{\antc}[1]   {\left\{ #1 \right\}}
\newcommand{\bra}[1]    {\langle #1 |}
\newcommand{\ket}[1]    {| #1 \rangle}
\newcommand{\eqn}[1]    {(\ref{#1})}
\newcommand{\subbox}[1] {{\mbox{\scriptsize #1}}}
\newcommand{\ex}[1] {{e^{ #1 }}}

\def\etal   {{\em et al.}}

\def\bi         {\begin{itemize}}
\def\ei         {\end{itemize}}
\def\benu   {\begin{enumerate}}
\def\eenu   {\end{enumerate}}
\def\bmat       {\left( \begin{array}}
\def\emat       {\end{array} \right)}
\def\beq    {\begin{equation}}
\def\eeq    {\end{equation}}
\def\beqn       {\begin{eqnarray*}}
\def\eeqn       {\end{eqnarray*}}
\def\beqa       {\begin{eqnarray}}
\def\eeqa       {\end{eqnarray}}
\def\bquote {\begin{quote}}
\def\equote {\end{quote}}
\def\f          {\frac}

\def\a          {\alpha}
\def\b          {\beta}
\def\c          {\chi}
\def\d          {\delta}
\def\e          {\epsilon}
\def\et         {\eta}
\def\g          {\gamma}
\def\k          {\kappa}
\def\l      {\lambda}
\def\m          {\mu}
\def\n          {\nu}
\def\s          {\sigma}
\def\t          {\tau}
\def\th         {\theta}
\def\ve     {\varepsilon}
\def\vph    {\varphi}
\def\w          {\omega}
\def\x          {\xi}
\def\z      {\zeta}

\def\A          {\Alpha}
\def\B          {\Beta}
\def\D          {\Delta}
\def\E          {\Epsilon}
\def\Et         {\Eta}
\def\G          {\Gamma}
\def\tG         {\tilde{\Gamma}}
\def\La          {\Lambda}
\def\M          {\Mu}
\def\Si          {\Sigma}
\def\Th         {\Theta}
\def\W          {\Omega}
\def\X          {\Xi}

\def\ba         {{\bf a}}
\def\bb         {{\bf b}}
\def\bd         {{\bf d}}
\def\bdf        {{\bf f}}
\def\be         {{\bf e}}
\def\bg         {{\bf g}}
\def\bh         {{\bf h}}
\def\bk         {{\bf k}}
\def\bl         {{\bf l}}
\def\bm         {{\bf m}}
\def\bn         {{\bf n}}
\def\bp         {{\bf p}}
\def\bq         {{\bf q}}
\def\br         {{\bf r}}
\def\bs         {{\bf s}}
\def\bv         {{\bf v}}
\def\bx         {{\bf x}}
\def\by         {{\bf y}}
\def\bz         {{\bf z}}
\def\bA         {{\bf A}}
\def\bB         {{\bf B}}
\def\bD         {{\bf D}}
\def\bF         {{\bf F}}
\def\bG         {{\bf G}}
\def\bH         {{\bf H}}
\def\bJ         {{\bf J}}
\def\bM         {{\bf M}}
\def\bQ         {{\bf Q}}
\def\bR         {{\bf R}}
\def\bS         {{\bf S}}
\def\bX         {{\bf X}}
\def\bal        {{\mbox{\boldmath$\a$}}}
\def\bsi    {{\mbox{\boldmath$\s$}}}
\def\bDel       {{\mbox{\boldmath$\Delta$}}}

\def\cA     {{\mathcal{A}}}
\def\cB         {{\mathcal{B}}}
\def\cD     {{\mathcal{D}}}
\def\cH     {{\mathcal{H}}}
\def\cF     {{\mathcal{F}}}
\def\cG     {{\mathcal{G}}}
\def\cK     {{\mathcal{K}}}
\def\cM         {{\mathcal{M}}}
\def\cN     {{\mathcal{N}}}
\def\cO     {{\mathcal{O}}}
\def\cP     {{\mathcal{P}}}
\def\cR     {{\mathcal{R}}}
\def\cV     {{\mathcal{V}}}

\def\hd     {{\hat{d}}}
\def\hk     {{\hat{k}}}
\def\hs     {{\hat{\s}}}
\def\hbr    {{\hat{\br}}}
\def\hbs    {{\mbox{\boldmath$\hat{\s}$}}}
\def\hbx    {{\hat{\bx}}}
\def\hby    {{\hat{\by}}}
\def\hbz    {{\hat{\bz}}}
\def\hps    {{\hat{\psi}}}
\def\hu     {{\hat{u}}}
\def\hv     {{\hat{v}}}
\def\hA     {{\hat{A}}}
\def\hC     {{\hat{C}}}
\def\hD     {{\hat{D}}}
\def\hG     {{\hat{G}}}
\def\hF     {{\hat{F}}}
\def\hH     {{\hat{H}}}
\def\hV     {{\hat{V}}}
\def\hO     {{\hat{O}}}
\def\hU     {{\hat{U}}}
\def\hcD    {{\hat{\mathcal{D}}}}
\def\hcF    {{\hat{\mathcal{F}}}}
\def\hcG    {{\hat{\mathcal{G}}}}
\def\hDel   {{\hat{\D}}}
\def\hSig   {{\hat{\Si}}}
\def\hGu    {{\hat{\underline{G}}}}
\def\hCk    {{\hat{C}_\bk^{}}}
\def\hCkd   {{\hat{C}_\bk^\dagger}}
\def\hCa    {{\hat{C}_\alpha^{}}}
\def\hCad   {{\hat{C}_\alpha^\dagger}}
\def\hCb    {{\hat{C}_\beta^{}}}
\def\hCbd   {{\hat{C}_\beta^\dagger}}
\def\ho     {{\hat{1}}}

\def\tila   {{\tilde{a}}}
\def\tilal  {{\tilde{\a}}}
\def\tilet  {{\tilde{\et}}}
\def\tilG   {{\tilde{G}}}
\def\tilK   {{\tilde{K}}}
\def\tilcG  {{\tilde{\cG}}}
\def\tilw   {{\tilde{\w}}}
\def\tilT   {{\tilde{T}}}
\def\tg     {{\tilde{\g}}}
\def\tD     {{\tilde\Delta}}
\def\bn     {{{\bf\nabla}}}
\def\tR     {{\tilde{R}}}
\def\trho     {{\tilde{\rho}}}

\def\ckD    {{\check{D}}}
\def\ckg    {{\check{g}}}
\def\ckSi   {{\check{\Sigma}}}
\def\oo     {{\otimes}}
\def\barK   {{\bar{K}}}
\def\barcG  {{\bar{\cG}}}

\def\vA     {{\vec{A}}}
\def\vB     {{\vec{B}}}
\def\vM     {{\vec{M}}}
\def\vk     {{\mathbf{k}}}
\def\kp     {{k_{\perp}}}

\def\la     {\langle}
\def\ra     {\rangle}
\def\larr   {\leftarrow}
\def\rarr   {\rightarrow}
\def\llarr  {\longleftarrow}
\def\Llarr  {\Longleftarrow}
\def\lrarr  {\longrightarrow}
\def\Lrarr  {\Longrightarrow}
\def\uparr  {\uparrow}
\def\dnarr  {\downarrow}
\def\dag    {\dagger}

\def\tr     {{\mbox{Tr}~}}
\def\im     {{\mbox{Im}}}
\def\re     {{\mbox{Re}}}
\def\Res    {{\mbox{Res}}}
\def\sgn    {{\mbox{sgn}~}}

\def\cm     {{\mbox{cm}}}
\def\kOe    {{\mbox{kOe}}}

\def\mca        {{M_{\mbox{\scriptsize AF}}}}
\def\mis        {{M_{\mbox{\scriptsize SP}}}}
\def\avmca      {{\bar{M}_{\mbox{\scriptsize AF}}}}
\def\avmis      {{\bar{M}_{\mbox{\scriptsize SP}}}}
\def\aca        {{\a_{\mbox{\scriptsize AF}}}}
\def\ais        {{\a_{\mbox{\scriptsize SP}}}}
\def\bca        {{\b_{\mbox{\scriptsize AF}}}}
\def\bis        {{\b_{\mbox{\scriptsize SP}}}}
\def\bas        {{\b_{\mbox{\scriptsize AF-SP}}}}
\def\ani        {{a_{\mbox{\scriptsize A}}}}
\def\bni        {{b_{\mbox{\scriptsize A}}}}
\def\abr        {{a_{\mbox{\scriptsize B}}}}
\def\bbr        {{b_{\mbox{\scriptsize B}}}}
\def\vphb       {{\vph_{\mbox{\scriptsize B}}}}

\def\vphni      {{\vph_{\mbox{\scriptsize A}}}}
\def\vphb       {{\vph_{\mbox{\scriptsize B}}}}
\def\mni        {{m_{\mbox{\scriptsize A}}}}
\def\mb         {{m_{\mbox{\scriptsize B}}}}
\def\xini       {{\xi_{\mbox{\scriptsize A}}}}
\def\xibr       {{\xi_{\mbox{\scriptsize B}}}}
\def\fni        {{f_{\mbox{\scriptsize A}}}}
\def\fbr        {{f_{\mbox{\scriptsize B}}}}

\def\Ck     {{C_\bk^{}}}
\def\Ckd    {{C_\bk^\dagger}}
\def\cks    {{c_{k {\sigma}}}}
\def\cku    {{c_{k {\uparrow}}}}
\def\ckd    {{c_{k {\downarrow}}}}
\def\cksd    {{c_{k {\sigma}}^\dagger}}
\def\ckud    {{c_{k {\uparrow}}^\dagger}}
\def\ckdd    {{c_{k {\downarrow}}^\dagger}}
\def\cmkdd    {{c_{{-k} {\downarrow}}^\dagger}}
\def\cmkd    {{c_{{-k} {\downarrow}}}}
\def\cmkpd    {{c_{{-k'} {\downarrow}}}}
\def\ckpu    {{c_{{k'} {\uparrow}}}}
\def\aks    {{a_{k {\sigma}}}}
\def\aku    {{a_{k {\uparrow}}}}
\def\akd    {{a_{k {\downarrow}}}}
\def\aksd    {{a_{k {\sigma}}^\dagger}}
\def\akud    {{a_{k {\uparrow}}^\dagger}}
\def\akdd    {{a_{k {\downarrow}}^\dagger}}

\def\uup    {{u_{{\uparrow}}}}
\def\udo   {{u_{{\downarrow}}}}
\def\vup    {{v_{{\uparrow}}}}
\def\vdo    {{v_{{\downarrow}}}}

\def\brag   {{\langle|}}
\def\ketg   {{|\rangle}}

\def\plx     {\f{\partial}{\partial x}}
\def\uu      {\uparrow\uparrow}
\def\ud      {\uparrow\downarrow}
\def\du      {\downarrow\uparrow}
\def\dd      {\downarrow\downarrow}